# Upper critical field, critical current density and thermally activated flux flow in CaFFe$_{0.9}$Co$_{0.1}$As superconductor

Chandra Shekhar[1,2,a)], Amit Srivastava[2], Pramod Kumar[3], Pankaj Srivastava[2], O. N. Srivastava[2]

[1]*Institut für Anorganische und Analytische Chemie, Johannes Gutenberg-Universität, 55099 Mainz, Germany*
[2]*Centre of Advanced Studies for Physics of Materials, Department of Physics, Banaras Hindu University, Varanasi-221005, India*
[3]*Department of Physics, Tulane University, New Orleans 70118 LA, USA*

[a)] Electronic mail: shekhar@uni-mainz.de, cgsbond@gmail.com

In this paper, we report the synthesis, structure, transition temperature, upper critical field, critical current density and thermally activated flux flow in CaFFe$_{0.9}$Co$_{0.1}$As superconductor. Superconductivity arises at 23 K by Co substitution at the site of Fe atoms and upper critical field is estimated 102 T by using the Werthamer-Helfand-Hohenberg formula. The flux-flow activation energy ($U_0/k_B$) varies from 3230 K and 4190 K in the field of 1 T and 9 T respectively. At 2K, the $J_c$ is found to be approximately $4\times10^3$ A/cm$^2$ and $0.3\times10^3$ A/cm$^2$ in zero and 6T field, respectively. Transmission electron microscopy analysis shows an amorphous region surrounding around most of the grains which is likely to be present in the form of amorphous and weak link grain boundaries in this compound. It seems that most of the current is hindered by miss aligned grains, amorphous grain boundaries and impurities, which are invariably found between the grains. Presence of the weakly linked granules and their weakly pinned intergranular Josephson vortices are responsible for both low $J_c$ and the Arrhenius temperature dependence of resistivity

# 1. Introduction

The discovery of the new family of iron arsenide, $LaO_{1-x}F_xFeAs$ (La-1111), superconductor [1] has stimulated considerable pursuit to scientific community in the area of the condensed matter especially high temperature superconductors and strongly correlated electron systems. The extensive research efforts have been devoted to this system due to the relative high transition temperature ($T_c$) in the presence of iron, layered structure and structural similarity to the cuprates. From replacement of tri-valence La by other rare-earth elements in ROFeAs (R: rare-earth elements), $T_c$ has been increased significantly and attained the record of 55K [2-6] at ambient pressure. Another family of the FeAs-based superconductor system has been discovered in $Ba_{1-x}K_xFe_2As_2$ (Ba-122) with $T_c$ at 38 K [7, 8]. The parent compounds for these superconductors belong to the ZrCuSiAs-type structure (space group P4/nmm) or $ThCr_2Si_2$-type structure (space group I4/mmm), consisting of an alternating stack of $(RO)^{+\delta}$ or $(Ba)^{+\delta}$ and $(FeAs)^{-\delta}$ layers. These compounds adopt the layered structure with a single FeAs layer per unit cell of ROFeAs and two such layers per unit cell of $BaFe_2As_2$. It is believed that FeAs layers are responsible for the superconductivity [9, 10]. This is quite similar to that of cuprates in which CuO layers are responsible for superconductivity as in YBCO compound. Another homologous series of FeAs containing compounds, AFFeAs (A = Ca, Sr and Eu) in which the $(FeAs)^{-\delta}$ layer is sandwiched by the $(AF)^{+\delta}$ layer in place of the $(RO)^{+\delta}$ layer in ROFeAs compounds. The AFFeAs compounds, like other pnictides, show the spin density wave (SDW) in between 120-180 K [11-15]. However, the superconductivity is induced by the partial substitution of some elements of lanthanide series [11, 16, 17] at A site and by the partial substitution of Co at Fe site [12-14, 18-21]. Following the partial substitution of Co at Fe site in pnictides leads to the highest observed $T_c$ of 14K for $LaOFe_{0.85}Co_{0.15}As$ [18], 17K for $SmOFe_{0.9}Co_{0.1}As$ [19], 20K for $SrFe_{1.8}Co_{0.2}As_2$ [20] and 22K for $BaFe_{1.8}Co_{0.2}As_2$ [21], so far. Further, highest $T_c$ is 4K for $SrFFe_{0.87}Co_{0.13}As$ [12] and 22K for 10% Co substituted CaFFeAs i.e. $CaFFe_{0.9}Co_{0.1}As$ [13]. Among the transition metal doping such as Co, Cr, Cu, Ir, Mn and Ni [14,22], only Co doping has invoked the effective emergence of superconductivity up to 22K and it also shows the superconducting transition for a wide range of Co concentration from 0.05 to 0.26 [14]. In light of these studies, the Co substitution is a potential way to convert FeAs-based layered compounds to superconductors.

For practical applications of a superconductor, two of the most important parameters are the upper critical field, $H_{c2}$, and the critical current density, $J_c$. The upper critical field is an intrinsic property, which has been approximated to be higher than 68 or 64T [23, 24] in $LaO_{0.9}F_{0.1}FeAs$, 70 T in $PrO_{0.85}F_{0.15}FeAs$, over 100 T in $SmFeAsO_{0.85}F_{0.15}$ [25], and 230 T in high-pressure fabricated $NdFeAsO_{0.82}F_{0.18}$ [26]. It is well known that the $J_c$ can be controlled by the flux pinning behaviour. However, critical current density, pinning force and upper critical fields of $CaFFe_{0.9}Co_{0.1}As$ compound are not reported so far. In the present paper, we report the study of the structures, microstructures and investigated the critical current density, and upper critical fields of $CaFFe_{0.9}Co_{0.1}As$ compound. In this connection, we have synthesized a series of compounds with different Co doping concentrations but dealt here only 10at. % Co doped sample, due to highest $T_c$ and larger superconducting volume fractions with the pure one .Our result shows that $J_c$ is quite sensitive to the temperature and also thermally activated flux flow is responsible for the broadening of the transition at $T_c$ in high magnetic field. An upper critical field for $CaFFe_{0.9}Co_{0.1}As$ is 102 T by using the Werthamer-Helfand-Hohenberg formula.

## 2. Experimental details

Polycrystalline samples with the nominal composition of CaFFe$_{1-x}$Co$_x$As (x= 0.0, 0.05, 0.07, 0.1 and 0.13) were prepared by conventional solid state reaction using high quality CaF$_2$, Fe, Co, and CaAs as the starting materials. Ca and As was grounded in 1:1 ratio and pressed into pellets, then sintered at 700$^0$C for 12 h in evacuated quartz tube. The final stoichiometric materials were taken according to 1/2CaF$_2$+CaAs+(1−x) Fe + x Co =CaFFe$_{1-x}$Co$_x$As and were heated at 1050$^0$C for 40 h containing different concentration of cobalt metal in evacuated sealed quartz tube. The X-ray diffraction of samples was performed with Cu-K$_\alpha$ radiation in the 2θ range from 5° to 80°, with a step interval of 0.01°. Microstructures of the as-obtained samples and their morphology were studied using environmental scanning electron microscopy (SEM, Quanta 200), operated at 30 kV. Microstructural characterization was carried out by high resolution transmission electron microscopy (HRTEM, FEI Tecnai20G$^2$, operated at 200 kV). Transport and magnetic properties were measured over a wide range of temperature and magnetic fields up to 7 T using a physical properties measurement system (PPMS, Quantum Design). Critical current density was calculated using the Bean model. Crystal structures were refined using Rietveld refinement.

## 3. Results and Discussion

### 3.1. Crystal Structure

The as-synthesized samples were subjected to the gross structural characterization employing the x-ray diffraction technique. Typical X-ray diffraction (XRD) pattern of CaFFeAs and CaFFe$_{0.9}$Co$_{0.1}$As is presented in figure 1. It can be seen that the samples are nearly single phasic. Rietveld refinement results confirm ZrCuSiAs type tetragonal structure with space group P4/nmm having lattice parameters $a$ = 3.877 Å, $c$ = 8.596 Å & $a$ = 3.884 Å, $c$ = 8.554 Å for pure and 10% Co doped samples respectively. These values are very close (little higher) to the reported standard lattice parameter values [19, 20] which indicate that Co atoms have been substituted in FeAs layers successfully.

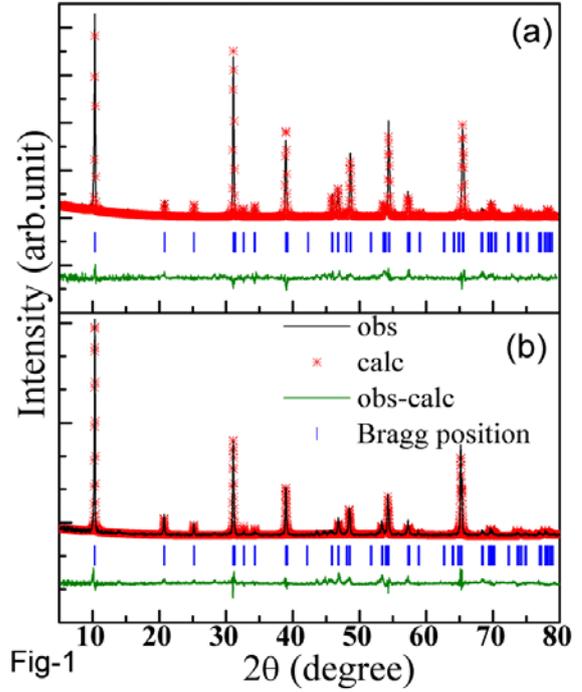

**Figure 1.** *Observed (solid line), calculated (symbols), Bragg position (vertical line) and difference (bottom line) XRD pattern of (a) CaFFe$_{0.9}$Co$_{0.1}$As (b)CaFFeAs*

### 3.2 Superconducting transition temperature

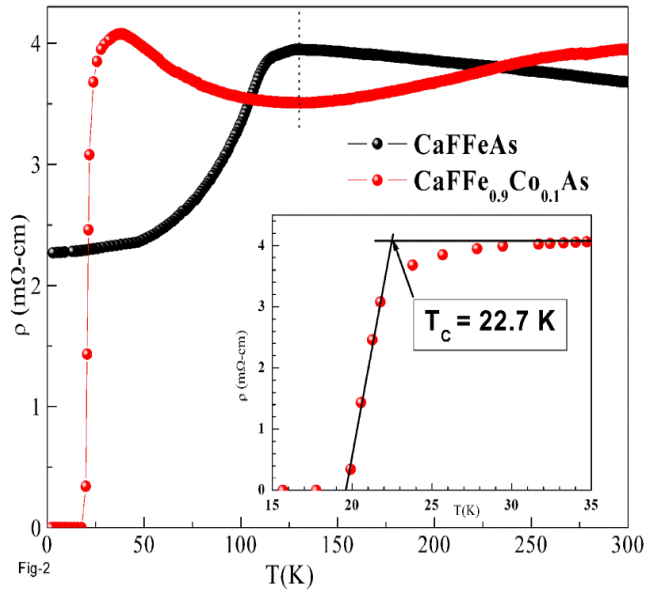

**Figure 2.** *Resistivity behaviour of CaFFeAs and CaFFe$_{0.9}$Co$_{0.1}$As with temperature, dotted line clearly shows the suppression of SDW in doped sample. Inset shows the enlarged view of transition point.*

The resistivity behaviour with temperature of pure and 10% Co-substituted CaFFeAs samples have been presented in figure 2. For CaFFeAs, the resistivity increases slightly with decrease in temperature, but below ~130 K, the resistivity drops steeply implying SDW anomaly to be at 130 K in CaFeAsF sample. This SDW relates the structural distortion followed by the anti-ferromagnetic order of Fe spins as revealed by neutron diffraction experiments [27]. However, 10% Co-substituted CaFeAsF sample showed $T_c$ at 22.7 K as shown in the in-set of figure 2. It is also pointed out that maxima point of pure and minima point of doped samples in the resistivity curves lie nearly at same temperature (see dotted line in figure 2). This shows that SDW is completely suppressed and superconductivity arises due to Co doping. The superconducting transition width, $\Delta T_c$, ($T_{c, on} - T_c$) is found to be nearly 2.3 K. In order to confirm further, the superconductivity and superconducting volume fraction, magnetic susceptibility ($\chi$) for CaFFe$_{0.9}$Co$_{0.1}$As sample were investigated at 5 and 20 mT in both zero field cooled (ZFC) and field cooled (FC) as shown in figure 3. The susceptibility becomes negative below 23 K as shown in the in-set of figure 3. The sharp diamagnetic superconducting transition from Meissner effect indicates good sample quality, and high superconducting shielding volume fraction reveals the bulk nature of this superconductor. Assuming the theoretical density of roughly 6.68 g/cm$^3$ for the perfect diamagnetism, we estimated the shielding and Meissner fractions, which are found to be about 90% and 2%, respectively at 2 K. It should be noted that since the Meissner fraction is determined by pinning and penetration effects, its interpretation is quite ambiguous on polycrystalline samples. The onset diamagnetic superconducting transition temperature is the same as that of the onset transition temperature on the corresponding resistivity curve. The resistivity and magnetization measurements demonstrate that 10% Co-substituted CaFeAsF is a bulk superconductor at~23 K. This is the highest achieved $T_c$ in Co doped 1111 pnictides compounds. Higher $T_c$ might be expected due to the correlation between $T_c$ and structure of FeAs tetrahedron. Recently, Zhao et al. [28] found that the highest $T_c$ of FeAs-based compounds can be obtained when the Fe-As-Fe bond angle is very close to the standard value of 109.47°, as expected for a perfect FeAs tetrahedron. The efficient way to increase $T_c$ in FeAs-based systems is to optimize the Fe-As-Fe angle. It can be noticed that Fe-As-Fe bond angle in CaFFeAs is 108.55° which is relatively close to the standard value than for the other pnictides [29-32]. Therefore, CaFFeAs can be considered as a promising system for maximizing $T_c$ by substitution.

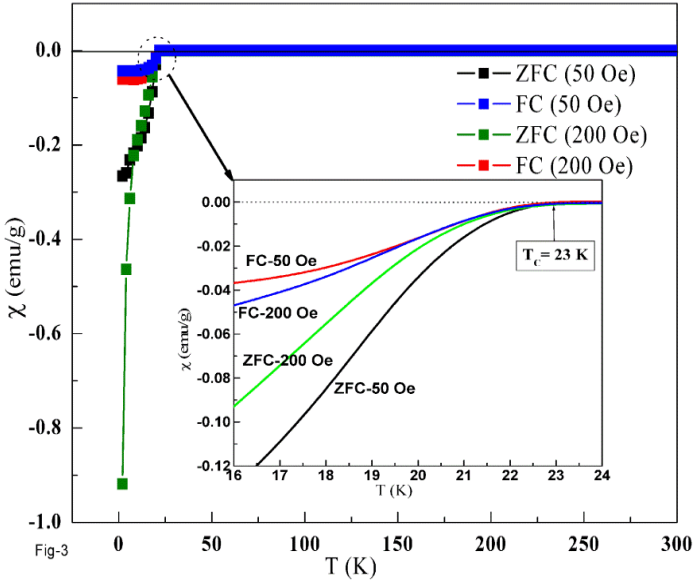

**Figure 3.** *Temperature dependence of the susceptibility measured in zero-field cooled (ZFC) and field cooled (FC, 50 and 2000 Oe) conditions. Inset shows enlarged view of transition point.*

*3.3. Upper critical fields*

Figure 4 presents the temperature dependence resistivity of 10% Co doped sample at different magnetic fields, B, up to 9 T. It can be seen that the onset $T_{c, on}$ drops very slow with increasing magnetic field. However, the $T_c$ shifts quickly to lower temperatures with the increase in magnetic field and the transition width becomes wider with increasing B. This is understandable in terms of flux creep in granular polycrystalline materials. The $T_c$ shifts quickly to lower temperatures with the increase in magnetic field is determined by the weak links between the grains as well as the vortex flow behavior, while $T_{c, on}$ drops very slow with increasing magnetic field is controlled by the upper critical field of the individual grains. This allows us to determine the upper critical field of this material. The upper critical field, $B_{c2}$, is defined as field at which the resistivity raises and approaches to normal state resistivity. Here, we have used a criterion of 90% of normal resistivity at the onset temperature and the $B_{c2}$ defined in this way refers to the case of a field parallel to the ab-plane, $B_{c2}^{ab}$. The 90% and 10% points of normal state resistivity can be regarded as the upper critical field parallel to $ab$ plane, $B_{c2}^{ab}$ and parallel to $c$ plane, $B_{c2}^{c}$ [24]. The values of the $B_{c2}^{ab}$ and $B_{c2}^{c}$, determined through the above fore said manner has now been plotted with temperature, are presented in the figure 5. The slope $dB_{c2}^{ab}/dT$ for 90% $\rho_n$ is estimated to be -6.45 T K$^{-1}$. This value is larger than

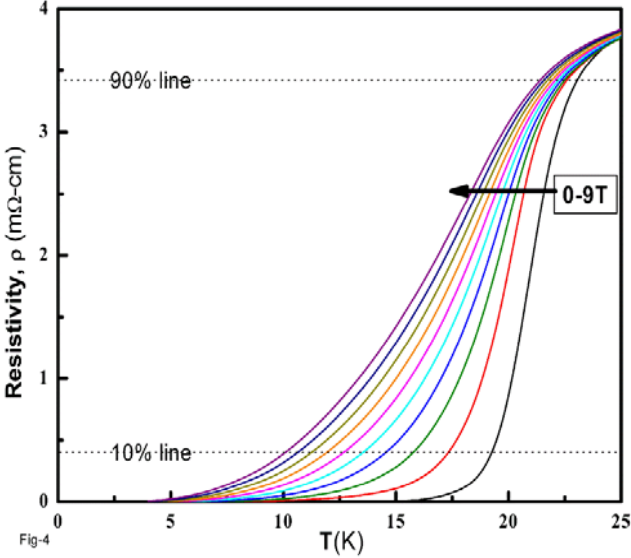

**Figure 4.** *Temperature dependence of resistivity under different magnetic fields. The onset transition point shifts weakly with the magnetic field. The dashed lines indicate the 10% and 90% points of the onset resistivity.*

that for La-1111 ($dB_{c2}/dT$ = -2 TK$^{-1}$) [23] and for Nd-11111 ($dB_{c2}/dT$ = -5.8 TK$^{-1}$) [25, 26]. Similarly the slope $dB_{c2}^{c}/dT$ for 10% $\rho_n$ is -1.6 T K$^{-1}$ at T≤ 13K. The $B_{c2}$ (0) can be estimated using single band Werthamer-Helfand-Hohenberg (WHH) model [33, 34] which is hardly applicable for iron based superconductors [24, 26] . This model is commonly used to evaluate the T = 0 limit of the $B_{c2}$ (0). Considering the orbital effects only into account (where the Maki parameter α is 0), the break-down of

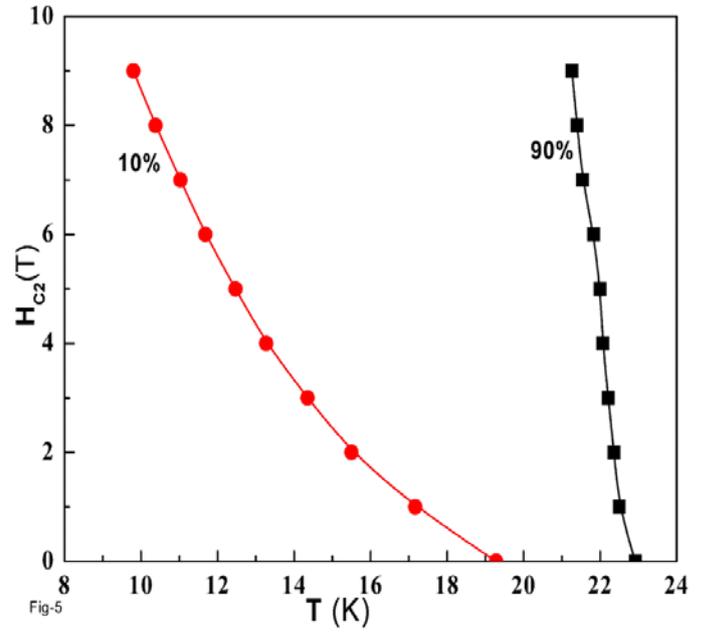

**Figure 5.** *Upper critical fields, $B_{c2}$ versus T as determined from the 10% and 90% points of the onset resistivity from figure 4.*

superconductivity predicted by the following formula: $B_{c2}^{ab}(0)$= -0.69 $T_c(dB_{c2}/dT)_{Tc}$ ,is 102 T for the 90% $\rho_n$ fields, and $B_{c2}^{c}(0)$ is 25 T for the 10% $\rho_n$ fields. Our result gives a rough estimate of $B_{c2}$ (0) because of the limit of the applied magnetic field. It may be noted that high values of $B_{c2}$ (0) can be achieved by: (i) strong band scattering, (ii) small Fermi velocities, and (iii) strong coupling. Strong coupling can be excluded empirically for Fe-As superconductors [35, 36]. It appears that the strong band scattering is responsible for high values of $B_{c2}$ (0). Lee et al. [37] have studied the behavior of the $B_{c2}$ and measured in field up-to 60

T in Sm-based pnictides which is determined by the complex interplay of a two-band nature and the Pauli paramagnetic effect depending on the direction of applied magnetic field with respect to the crystal axes. However, $B_{c2}$ shows only Pauli paramagnetic behaviour in As-deficient La-based pnictides [38]. Finally, we can say that the WHH approximation could not be simply applied in this material.

*3.4. Thermally activated flux flow*

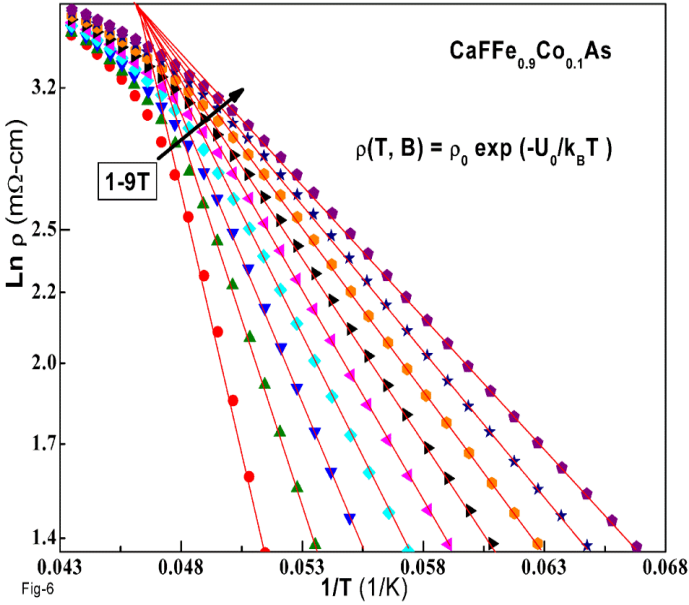

**Figure 6.** *Arrhenius plot of the resistivity. The activation energy $U_0$ at a field is given by the slope from linear fitting.*

The broadening of the ρ (T) (just above the $T_c$) in magnetic field for superconductors is interpreted in terms of a dissipation of energy caused by the motion of vortices [39-41]. This interpretation is based on the fact that, for the low-resistivity region, the resistivity is caused by the creep of vortices so that the ρ(T) dependences are thermally activated and are usually described by an Arrhenius equation

$$\rho(B,T) = \rho_0 \, exp\left(-\frac{U_0}{k_B T}\right)$$

where $U_0$ is the thermally activation flux-flow (TAFF) energy, which can be obtained from the slope of the linear part of an Arrhenius plot, $\rho_0$ is a field-independent pre-exponential factor, and $k_B$ is Boltzmann's constant. The best fitted ln ρ vs. $T^{-1}$ plot to the experimental data yields the values of the activation energy ranging down from $U_0/k_B = 3230$ K and 4190 K in the field of 1 T and 9 T, respectively. The flux-flow activation energy generally varies from 3000 to 300K with field from 1-9 T in BiSrCaCuO system [42] and in the case of $MgB_2$, it is around 10000K in field ≤1T and down to 300K in field of 10T [40], Since, the resistivity measurements of the superconducting transition of the different superconducting materials give insight into the flux pinning properties so that TAFF differ for different materials. Further, the power law field dependence of the activation energy $U_0$ (B) α $B^{-n}$ with the exponent n≤1 which usually observed for other layered systems [39-45]. Figure 7 presents the magnetic field dependence of the activation energy $U_0$ of $CaFF_{0.9}Co_{0.1}FeAs$. We can see that the values of $U_0$ for $CaFF_{0.9}Co_{0.1}FeAs$ drop weakly with field for B ≤3 T, scaled as $B^{-0.05}$, and then decrease as $B^{-0.15}$ for B>3T. In order to get broad insight of variation of $U_0$ is basically due to the pinning of the flux line. However, it is found that n= 1/3 for B < 3T and n = 1/4 for B > 3 T in BaSrCaCuO [42] and n = 1/6 for B < 2T and n = 1/3 for B > 2 T in $MgB_2$ film [40]. Relatively slow decrease of U0 (B) in low fields implies that weakly pinned intergranular Josephson vortices dominate (single vortex pinning) [46] followed by a quick decrease of U0 (B) in field, which could be related to a crossover to a collective flux creep regime [47]. This dissipative phenomenon is associated with the vortex motion inside the material.

*3.5 Critical current density*

Magnetization hysteresis loops (M-H loop) at 5, 7, 15K and 25K are measured by superconducting quantum interference device magnetometry (SQUID) as shown in figure 8. However, M-H loop at 25 K (above $T_c$) is measured to know the background. In-set shows the paramagnetic background, due to small amount of impurity phases, similar to earlier studies [1] at 25K. After the subtraction of background, the complete hysteresis loops are shown in figure 8. The hysteresis loops are combination of two different contributions, $M_{eq}$ or $M_{rev}$ and $M_{irr}$ where, $M_{eq}$ or $M_{rev}$ and $M_{irr}$ are equilibrium or reversible and irreversible magnetizations respectively. Here, the values of $M_{eq}$ and $M_{irr}$ at 2K are 0.97emu/g and 1.13enu/g respectively. The sample showed a high $B_{c2}^{ab}$ (5K) of 102 T, only small hysteresis loops were found as compared to other polycrystalline iron oxypnictides [48, 49]. These small hysteresis loops of present sample reveal either weak flux pinning and/or weak inter-granular coupling. However, strong inter-granular coupling implies very high hysteresis loop in the case of $MgB_2$ [50]. The magnetic field dependence of the critical current density $J_c$ derived from the irreversible parts of the hysteresis loop width using the extended Bean's model

$$J_c = \frac{20 \times \Delta M}{Va(1 - a/3b)}, \qquad a < b$$

Where, $\Delta M$ is the width of the hysteresis loop measured in emu, V is volume of sample in $cm^3$, and a, b are respective sample dimensions in cm, $J_c$ is in $Acm^{-2}$. For estimation of $J_c$, the full sample dimensions of $2.0 \times 1.0 \times 0.5$ $mm^3$ were taken. The $J_c$ vs. magnetic field at different temperatures extracted from the hysteresis loop widths using the Bean model is shown in figure 9. At 2 K, the $J_C$ is approximately $4 \times 10^3$ A/$cm^2$ at zero field and then decreases to $1.5 \times 10^3$ A/$cm^2$ at 0.3 T. The $J_c$ increases slightly with increase in magnetic field above 0.6 T to a maximum of $1.6 \times 10^3$ A/$cm^2$. This peak effect has also been observed in $SmO_{1-x}F_xFeAs$ [48] and $CeO_{1-x}F_xFeAs$ [51] compounds. It is more clearly visible in figure 9. It can be observed from figure 9 that $J_c$ at 2 K rapidly decreases up to field 0.2 T. This indicates that impurities grain switch off when small magnetic field is applied and behave like Josephson junctions [52].

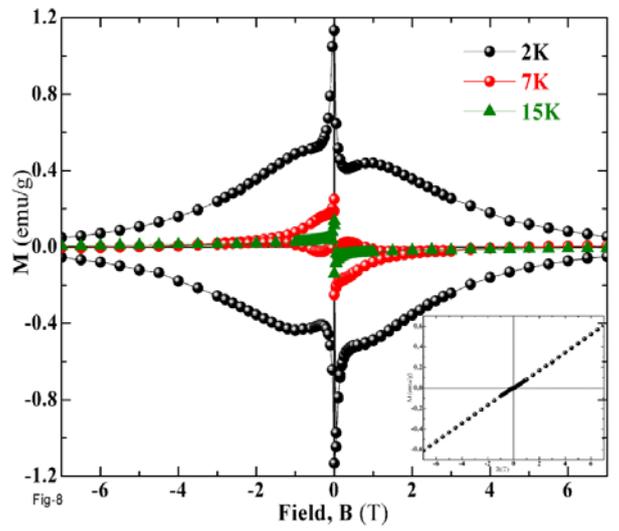

**Figure 8.** *Superconducting contribution to the magnetic moment in hysteresis loops as a function of field at 2, 5 and 7K. The superconducting signal was isolated from the paramagnetic background determined above $T_c$ i.e. 25K. Inset shows the paramagnetic background determined above $T_c$ i.e. 25K.*

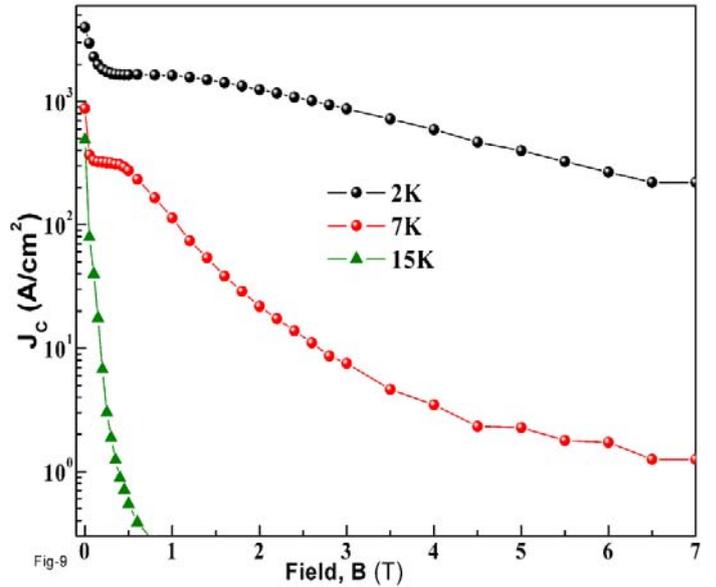

**Figure 9.** *Magnetic field dependence of the critical current density at different temperatures.*

### 3.6 Microstructures

The $J_c$ is one of the important parameters which is used to characterize the technologically important of the polycrystalline superconductors. The grain boundaries in the samples is assumed to play crucial role in determining the $J_c$. Therefore, in order to explore the microstructural characteristics and its possible correlation with superconducting properties, we have carried out SEM and TEM studies of the Co doped CaFFeAs sample. The extensive microstructural investigations reveal the following special characteristics as

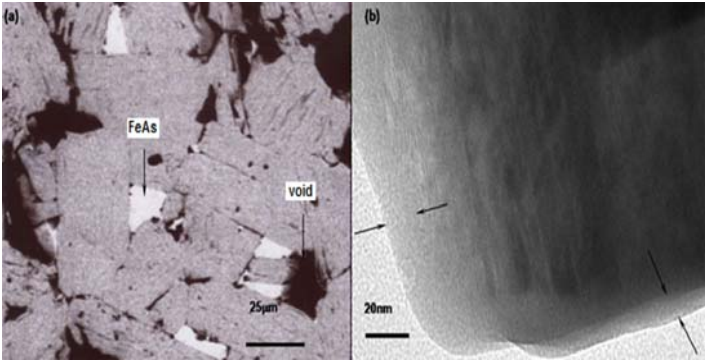

**Figure 10** (a) *Back scattered SEM image showing FeAs impurity phase and voids which are indicated by arrows (b) TEM image showing a thick amorphous region is surrounded around the grain which might be responsible for forming thick grain boundary between two grains.*

shown in figure 10 and figure 10(a) shows back scattered SEM of polished surface of the Co doped CaFFeAs sample. This sample consists of lager grains having size in the range of 15–40 $\mu m$, and has been identified as Co doped CaFFeAs grains. These grains are much larger than those reported previously [52-54]. However, some FeAs is also visible at some points which are marked by arrows in figure 10(a). These FeAs may be present sometimes between the boundaries of the grains and thus, these interrupt grain to grain super current paths. Further, study on the microstructure of the samples has also been performed using TEM. Figure 10(b) shows a typical grain of the Co doped CaFFeAs sample. Numerous randomly selected grains have been examined in this approach and during this course; it has been observed that there is an amorphous layer present around most of the grains in this sample. A typical example of this observation is presented in figure 10(b). A Detailed observation of a single grain has been performed and an amorphous layer of several nanometers [a region between the arrows in figure 10(a)] in thickness around individual grains has been reflected. Evidence of a similar amorphous layer was also observed in polycrystalline $Sr_{0.6}K_{0.4}Fe_2As_2$ and $YBa_2Cu_3O_{7-\delta}$ [54, 55]. Generally, this amorphous layer forms thick and weak grain boundaries (GBs) which hinder most of the current.

Finally, we are tried to summarize the correlation of transport properties with microstructures. The transport properties are governed by grain boundaries. As we discussed above as well as some recent reports also suggested that the most of the GBs of iron based superconductors are weak linked [56, 57] similar way to GBs in the cuprate superconductors. Presence of weakly linked granules and their weakly pinned intergranular Josephson vortices are responsible for the TAFF resistivity, which can further responsible for both low $J_c$ and the Arrhenius temperature dependence of resistivity. For further improvement in $J_c$, it needs to modify the synthesis condition and tailoring of materials by some doping/admixing where intergranular Josephson vortices can easily be pinned.

## 4. Conclusion

Based on the above results and discussion it can be concluded that the transition temperature of 23 K in $CaFFe_{0.9}Co_{0.1}As$ compound is highest as compared to any transition elements doped 1111compound. The upper critical field is 102 T. At 2K, the $J_c$ is evaluated to be approximately $4 \times 10^3 A/cm^2$ and $0.3 \times 10^3 A/cm^2$ in zero field and 6T respectively. Therefore, it needs to further improvement of $J_c$ and $H_{c2}$ by improving the gains connectivity and creating disorder in this compound respectively.

## Acknowledgements

One of the authors (Chandra Shekhar) is grateful to Prof. Dr. B. Büchner for inviting as a guest scientist at IFW-Dresden, Germany. The financial supports from DST-UNANST and CSIR are gratefully acknowledged. AS is grateful to UGC for providing Teacher Fellowship under FIP Scheme. Author (Chandra Shekhar) is also grateful to UGC for the award of Dr D S Kothari postdoctoral fellowship.